\documentclass[twocolumn]{aastex631}
\usepackage{graphicx}
\usepackage{amsmath}
\usepackage{tablefootnote}
\usepackage{amsmath}
\usepackage{threeparttable}

\usepackage{natbib}

\newcommand{\black}{\textcolor{black}}

\begin{document}

\title{Measurement of the Forward Shock Velocities of the Supernova Remnant N132D \\ Based on the Thermal X-ray Emission }

\author[0009-0001-9828-0551]{Yoshizumi Okada}
\affiliation{Department of Science and Engineering, Graduate School of Science and Engineering, Aoyama Gakuin University, 5-10-1, Fuchinobe, Sagamihara 252-5258, Japan}
\affiliation{Institute of Space and Astronautical Science (ISAS), Japan Aerospace Exploration Agency (JAXA), 3-1-1 Yoshinodai, Chuo-ku,
Sagamihara, Kanagawa 252-5210, Japan}
\author[0009-0002-4783-3395]{Yuken Ohshiro}
\affiliation{Institute of Space and Astronautical Science (ISAS), Japan Aerospace Exploration Agency (JAXA), 3-1-1 Yoshinodai, Chuo-ku,
Sagamihara, Kanagawa 252-5210, Japan}
\affiliation{Department of Physics, Graduate School of Science, The University of Tokyo, 7-3-1 Hongo, Bunkyo-ku, Tokyo 113-0033, Japan}
\author[0009-0008-1853-6379]{Shunsuke Suzuki}
\affiliation{Department of Science and Engineering, Graduate School of Science and Engineering, Aoyama Gakuin University, 5-10-1, Fuchinobe, Sagamihara 252-5258, Japan}
\affiliation{Institute of Space and Astronautical Science (ISAS), Japan Aerospace Exploration Agency (JAXA), 3-1-1 Yoshinodai, Chuo-ku,
Sagamihara, Kanagawa 252-5210, Japan}
\author[0000-0002-8152-6172]{Hiromasa Suzuki}
\affiliation{Institute of Space and Astronautical Science (ISAS), Japan Aerospace Exploration Agency (JAXA), 3-1-1 Yoshinodai, Chuo-ku,
Sagamihara, Kanagawa 252-5210, Japan}
\author[0000-0003-1415-5823]{Paul P. Plucinsky}
\affiliation{Center for Astrophysics | Harvard \& Smithsonian, 60 Garden St., Cambridge, MA 02140, USA}
\author[0000-0002-1251-7889]{Ryo Yamazaki}
\affiliation{Department of Physical Sciences, Aoyama Gakuin University, 5-10-1, Fuchinobe, Sagamihara 252-5258, Japan}
\affiliation{Institute of Laser Engineering, Osaka University, 2-6 Yamadaoka, Suita, Osaka 565-0871, Japan}
\author[0000-0002-5092-6085]{Hiroya Yamaguchi}
\affiliation{Institute of Space and Astronautical Science (ISAS), Japan Aerospace Exploration Agency (JAXA), 3-1-1 Yoshinodai, Chuo-ku,
Sagamihara, Kanagawa 252-5210, Japan}

\begin{abstract}
Measuring shock velocities is crucial for understanding the energy transfer processes at the shock fronts of supernova remnants (SNRs), including acceleration of cosmic rays. 
Here we present shock velocity measurements on the SNR N132D, based on the thermal properties of the shock-heated interstellar medium.
We apply a self-consistent model developed in our previous work to X-ray data from deep Chandra observations with an effective exposure of $\sim$\,900\,ks.  
In our model, both temperature and ionization relaxation processes in post-shock plasmas are simultaneously calculated, so that we can trace back to the initial condition of the shock-heated plasma to constrain the shock velocity. 
We reveal that the shock velocity ranges from 800 to 1500 $\rm{km~s^{-1}}$ with moderate azimuthal dependence. 
Although our measurement is consistent with the velocity determined by independent proper motion measurements in the south rim regions, a large discrepancy between the two measurements (up to a factor of 4) is found in the north rim regions. 
This implies that a substantial amount of the kinetic energy has been transferred to the nonthermal component through highly efficient particle acceleration. 
Our results are qualitatively consistent with the $\gamma$-ray observations of this SNR.

\end{abstract}


\section{Introduction} \label{sec:intro}

It is widely believed that supernova remnants (SNRs) are the major source of Galactic cosmic-rays with energies up to $\gtrsim 100$\,TeV, through the process called Fermi acceleration \citep[e.g.,][]{1934PhRv...46...76B,1978MNRAS.182..147B,1978ApJ...221L..29B,1995Natur.378..255K}. 
A velocity of the SNR shock wave is thought to be one of the key parameters that determine the cosmic-ray acceleration efficiency, and thus it is important to constrain it observationally. 
The most direct way to do this is to measure the proper motion of the shock front. To date, this method has been applied to a number of SNRs in the Milky Way and nearby galaxies, including SN1006 \citep{2009ApJ...692L.105K}, RCW 86 \citep{2016ApJ...820L...3Y}, \black{Tycho \citep{2021ApJ...906L...3T}} and Cassiopeia A \citep{Vink_2022}. 

In principle, thermal X-ray emission from the shock-heated intersteller medium (ISM) also provides information on the forward shock velocity. 
During the shock transition in low-density environments ($n \lesssim 1~\rm{cm}^{-3}$), such as those observed in SNRs, thermal equilibration among different particle species is not achieved immediately \citep[e.g.,][]{Rakowski_2003, Ghavamian_2007, Yamaguchi_2014}. 
Assuming the Rankine-Hugoniot shock jump conditions for monoatomic gas (i.e., $\gamma = 5/3$, where $\gamma$ is the adiabatic index), the postshock temperature of species $i$ is given as
\begin{equation}\label{eq:RH}
    kT_{\rm{i}}=\frac{1}{\chi} \left(1-\frac{1}{\chi}\right) m_{\rm i} v_{\rm sh}^2,
\end{equation}
where $k$, $\chi$, $m_{\rm{i}}$ and $v_\mathrm{sh}$ are the Boltzmann constant, compression ratio, the mass of the species $i$ and the shock velocity, respectively, 
in the case without collisionless energy transfer
among different species \citep{vink_relation_2010}.
The downstream plasma then gradually approaches thermal equilibrium via Coulomb collisions, with which collisional ionization proceeds simultaneously. 
Thermal emission from shocked plasma is mainly characterized by the electron temperature $kT_{\rm e}$ and ionization degree $\tau$ (= $\int_0^{t} n_{\rm e}~dt'$, where $n_{\rm e}$ and $t$ are the electron number density and elapsed time). Therefore, the immediate postshock electron temperature can be estimated by tracing back the thermal equilibration and the ionization processes to $t = 0$, and thus the shock velocity is determined from Equation~\ref{eq:RH}.

Recently, we have developed a self-consistent model of thermal X-ray emission from shock-heated plasmas, where the postshock processes of the temperature and ionization relaxation are simultaneously calculated: Ionization and Temperature Non-equilibrium Plasma (\texttt{IONTENP}) model
\citep{Ohshiro_2024}. 
This model enables us to estimate the shock velocity based on the thermal properties of the plasma constrained through X-ray spectroscopy, even for distant SNRs, where proper motion measurement is challenging. 
Moreover, by comparing with the proper motion velocity, we can investigate how much shock kinetic energy is transferred to nonthermal energy, in a similar way to the discussion in \cite{2000ApJ...543L..61H} and \cite{helder_measuring_2009}.

As the first application of the shock-velocity estimate based on the {\tt IONTENP} model, we focus on the SNR N132D, located in the Large Magellanic Cloud, the distance to which is approximately 50~kpc \citep{2019arXiv191012754K}.
The progenitor of this SNR is thought to be a massive star that exploded as a Type~Ib supernova \citep{2000ApJ...537..667B}. 
The age of N132D is estimated to be $2770\pm500$~years by measuring the proper motion of the oxygen-rich ejecta with a baseline of over 16 years with Hubble Space Telescope (HST) \citep{banovetz_hst_2023}.
This SNR is either in or entering the Sedov-Taylor phase \citep{1997A&A...324L..45F}. \cite{chen_supernova_2003} suggested that the forward shock in the south rim has been decelerated due to its impact on a cavity wall, which was likely formed by the stellar wind of the progenitor. This agrees with the distribution of the molecular and atomic gas on the periphery \citep{kim_neutral_2003, sano_alma_2020}.
More recently, \cite{2024AAS...24311312P} measured the proper motion of the forward shock using the Chandra X-ray Observatory.
They reported
the blast wave velocities of $1700\pm 400~\rm{km~s^{-1}}$ and $3700\pm500~\rm{km~s^{-1}}$ in the southern and the northern rims, respectively. 
Notably, N132D is the most luminous SNR in X-ray and $\gamma$-ray in the Large Magellanic Cloud (e.g., \citealp{Maggi_2016, 2021A&A...655A...7H}), while its synchrotron X-ray emission has not been firmly detected \citep{bamba_transition_2018}.
The dominant contributor to the $\gamma$-ray emission is likely the hadronic process. This suggests a possibility of highly efficient cosmic-ray acceleration (\citealp{sano_alma_2020,2021A&A...655A...7H}).

In this paper, we present our study to estimate the forward shock velocities of the SNR N132D by applying the \texttt{IONTENP} model to its thermal X-ray spectra. 
We describe the observation and data reduction in Section \ref{sec:data reduction}. Spectral analysis is presented in Section \ref{sec:Analysis}. We discuss the results in Section \ref{sec:discussion}. In Section \ref{sec:conclusion}, we summarize our results and interpretation.
Errors presented in the text, figures, and tables are at a 1$\sigma$ confidence level.

\section{Observation and Data Reduction} \label{sec:data reduction}

N132D was observed for an effective exposure of 868~ks in total with the Chandra X-ray Observatory from 2019 March 27 to 2020 July 16 (Table \ref{table:observation}). We use all the data since no significant background flares were detected.
The data were acquired using the S3 chip of the Advanced CCD Imaging Spectrometer (ACIS; \citealp{1997AAS...190.3404G}) detector array. 
We reprocessed the data using CIAO version 4.14 (\citealp{10.1117/12.671760}) and CALDB version 4.9.8 \citep{2007ChNew..14...33G} following the standard procedure using $\texttt{chandra\_repro}$. 
We use HEAsoft version 6.30.1 (HEASARC \citeyear{2014ascl.soft08004N}) and \texttt{XSPEC} version 12.12.1 \citep{1996ASPC..101...17A} to perform spectroscopy.

\section{Analysis and results} \label{sec:Analysis} 

\begin{figure} [ht]
\begin{center}
\includegraphics[width=80mm]{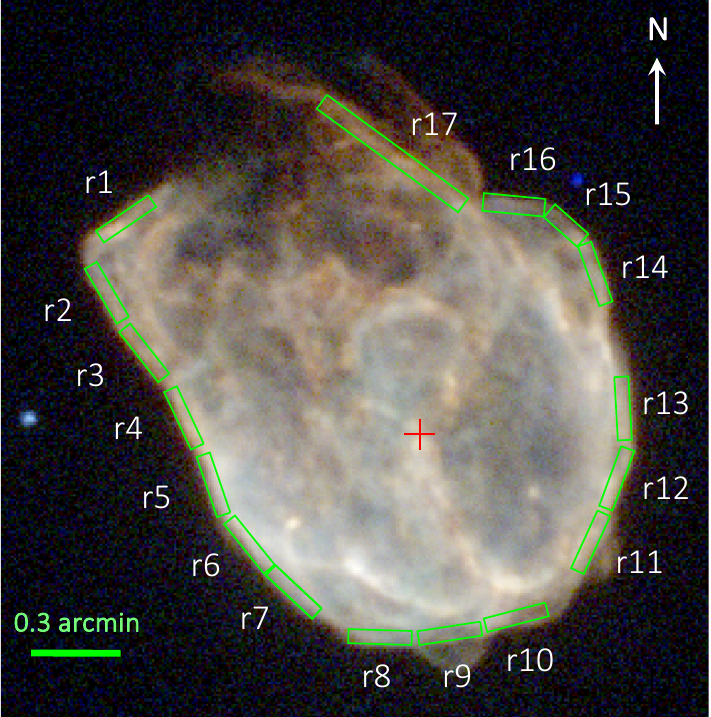}
\end{center} 
\caption{Chandra ACIS-S three color image (red:0.5--1.2~keV, green:1.2--2.0~keV, blue:2.0--7.0~keV)
with analysis regions (green boxes). The red cross indicates the inferred center of explosion \citep{banovetz_hst_2023}.}
\label{fig:region}
\end{figure}

We reproject all the observations and create a merged event file to improve the photon statistics using the tool \texttt{reproject\_obs}.
Figure \ref{fig:region} shows a three-color composite image of the soft (0.5--1.2~keV), medium (1.2--2.0~keV) and hard (2.0--7.0~keV) energy bands with the regions used for our analysis.
In order to extract the emission from the forward-shocked ISM, we use rectangular regions with their major axes parallel to the outer X-ray rim.
The background region for the spectral analysis is defined as a rectangle of $1\farcm2 \times 0\farcm7$ located $ 0\farcm9$ away from the center of N132D (Figure  
\ref{fig:bkg}). 
\color{black}
This background region is on the S3 chip for all the observations. 
\color{black}
We first determine a spectral model for the background region, and then use it in the modeling for the source spectra.
The C-statistic \citep{1979ApJ...228..939C} is employed to evaluate the spectral fit for unbinned spectra. 
Elemental abundances referred to in the text and tables are relative to the solar abundances given by \cite{ANDERS1989197}.

\subsection{Background Models} \label{ss:bkg}

\begin{figure} [h]
\begin{center}
\includegraphics[width=90mm]{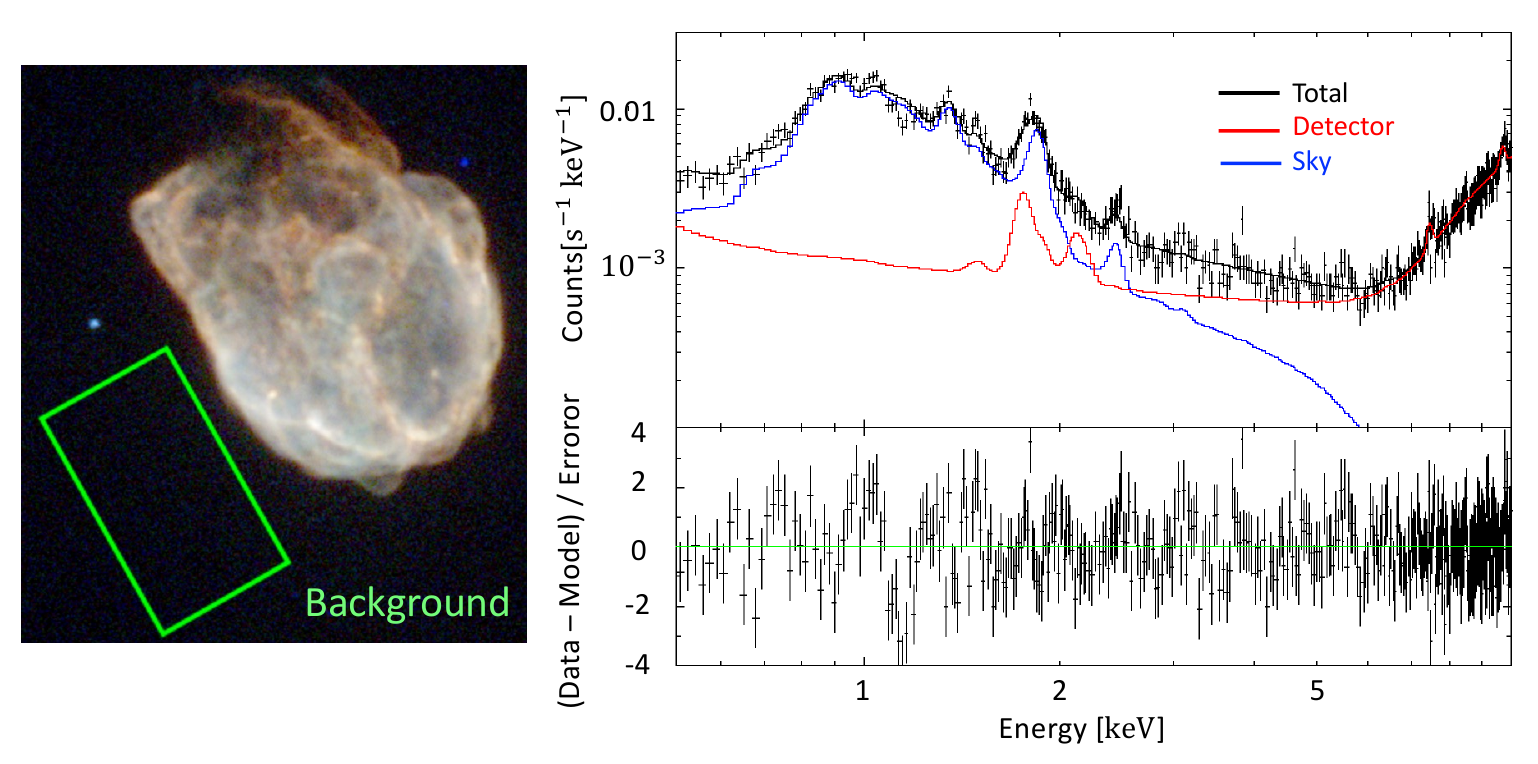}
\end{center} 
\caption{{\it (left)} Chandra three-color image and the background region (green box). {\it (right)} The background spectrum and best-fit model for background region. The background model consists of detector (red) and sky (blue) components. The black line indicates the entire model.}
\label{fig:bkg}
\end{figure}

In general, the background spectra can be divided into two components: the detector (non X-ray) background and the sky background. 
In order to model the detector background, we apply a tool, \texttt{mkacispback} \citep{suzuki_spatial_2021} to the merged data set.
Following \cite{sharda_spatially_2020}, we assume that the sky background is composed of Local Hot Bubble (\texttt{apec}, \citealp{Snowden_1997, kuntz_x-ray_2001, mccammon_high_2002, 2019arXiv191012754K}), Milkey Way Halo (\texttt{apec}, \citealp{Snowden_1998, snowden_catalog_2008}) and Cosmic X-ray Background (\texttt{powerlaw}) for background region.
The column density of the foreground absorption is treated as a free parameter (\texttt{tbabs}). 
The electron temperatures and normalization of the three components are also treated as free parameters, whereas the photon index of the {\tt powerlaw} is fixed to 1.46 \citep{Snowden_1997, 2004xmmg.rept.....S, Kuntz_2010}. The metal abundances are fixed to the solar values.
The spectral model for the background region reproduces the data well with the best-fit parameters shown in Table \ref{tab:sky} and with acceptable $\mathrm{C\text{-}stat\text{/}d.o.f.} =763\text{/}649$.
\color{black}
We do not engage in a detailed interpretation of each parameter because our purpose here is to construct an empirical background model.
\color{black}
The best-fit model and data are shown in Figure~\ref{fig:bkg}.

\begin{table} [ht] 
\centering \caption{Best-fit spectral parameters for the background region}
\begin{threeparttable}
    \begin{tabular}{c c c}
    \hline\hline
        Model & Parameter & Value \\ \hline
        Absorption & $\rm{N_{\rm H}}$~($\mathrm{10^{22}\,cm^{-2}}$) & $0.60_{-0.08}^{+0.05}$ \\
        LHB & $kT$~(keV) & $0.10_{-0.01}^{+0.01}$ \\ 
        ~ & Abundance & 1.00(fixed) \\ 
        ~ & Normalization\tnote{a} &  $4.3_{-2.8}^{+3.7} \times 10^{-2}$ \\ 
        MWH & $kT$~(keV) &  $0.56_{-0.02}^{+0.04}$\\ 
        ~ & Abundance & 1.00(fixed)  \\ 
        ~ & Normalization\tnote{a} & $1.6_{-0.3}^{+0.2}\times 10^{-4}$ \\ 
        CXB & Photon Index & 1.46(fixed) \\ 
        ~ & Normalization\tnote{b} & $6.3_{-0.5}^{+0.5} \times 10^{-6}$\\ \hline
    \end{tabular} 
    \begin{tablenotes}
    \item[a] Emission measure, $10^{-14}(4\pi D^2)^{-1}\int{n_{\rm e}n_{\rm H}dV}$, where $D$, $n_{\rm e}$, and $n_{\rm H}$ stand for distance (cm) and electron and hydrogen number densities ($\rm{cm^{-3}}$), respectively.
    \item[b]Normalization of the power-law model in units of $\rm{cm^{-2}\,s^{-1}\,keV^{-1}\,at\,1~keV}.$
    \end{tablenotes}
    \end{threeparttable}
     \label{tab:sky}
\end{table}

\subsection{Source Models} \label{ss:source} 

In order to model the source emission, we introduce the \texttt{IONTENP} model \citep{Ohshiro_2024}, which describes emission from a thermal plasma in the temperature and inoization non-equilibrium. 
We take into account two absorption components, the Galactic (\texttt{tbabs}) and LMC (\texttt{tbnew}) absorptions for the source emission in the 0.5–10.0~keV band.
We fix the column density of the Galactic absorption to $6.2\times\mathrm{10^{20
}\,cm^{-2}}$ \citep{1990ARA&A..28..215D}, while that of the LMC is set to be free.
The parameters of this model are elemental abundances, shock velocity ($v_\mathrm{sh}$), electron-to-proton temperature ratio just behind the shock ($\beta$), and ionization timescale ($\tau$).
The parameter $\beta$ reflects the efficiency of so-called collisionless electron heating, i.e., an energy transfer from protons to electrons in the shock transition layer.
Therefore, the allowable range of the $\beta$ value is 
$m_{\rm e}/m_{\rm p}\leq~\beta~\leq1.0$.  
$\beta=m_{\rm e}/m_{\rm p}$ means that no energy transfer takes place, so that the exact immediate postshock condition given in Equation~1 is kept. $\beta=1.0$ means that the plasma achieves the temperature equilibrium quickly.
Plasma states are insensitive  to $\beta$ when the ionization timescale $\tau$ $\gtrsim \rm{10^{11}~s ~cm^{-3}}$, as is the case for N132D (see Figure 7 of \citealp{Ohshiro_2024}).
In our analysis, therefore, $\beta$ is fixed to $m_{\rm e}/m_{\rm p}$.
The abundances of O, Ne, Mg, Si, S and Fe are free to vary. We fix the metal abundances to the values of LMC \citep{suzuki_plasma_2020} or to 0.3 for the elements without available values in the literature.
The emission from the plasma in the rim regions is explained by two components with different electron temperatures \citep{sharda_spatially_2020}, which may result from inhomogeneity in the ISM density ($n_{\rm e}$), leading to the difference in $\tau$.
Similarly, in our analysis, certain regions require two temperature models to explain the spectra in the energies below 1\,keV (O, Ne, Fe lines).
We use the spectral model {\tt vapec + IONTENP} for the source regions.
Here we note that a fit with {\tt vnei} instead of {\tt vapec} results in the ionization timescales of $\tau \gtrsim 10^{12}$~cm$^{-3}$~s, consistent with the ionization equilibrium.
The abundances of \texttt{vapec} model are linked with those of \texttt{IONTENP} model.
While the hard X-ray components including the Fe He$\alpha$ and Ly$\alpha$ emission are reported, they lie in the inner part of N132D likely because of their ejecta origin \citep{sharda_spatially_2020, collaboration_xrism_2024}. Thus, their contribution to our analysis regions should be negligible.

\begin{figure} [htb]
\begin{center}
\includegraphics[width=80mm]{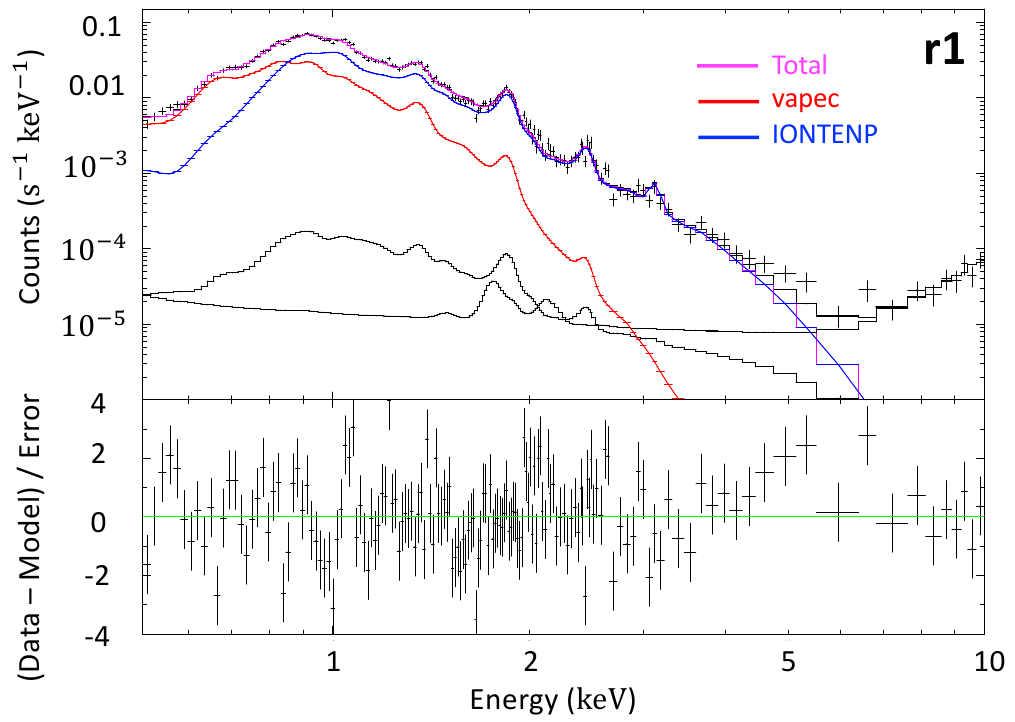}
\end{center} 
\caption{The X-ray spectrum and best-fit model for the region r1. The red, blue, black, and magenta solid lines represent the contributions of \texttt{vapec}, \texttt{IONTENP}, background (detector + sky) and total models, respectively.}
\label{fig:a}
\end{figure}

\begin{figure*} 
\begin{center}
\includegraphics[width=160mm]{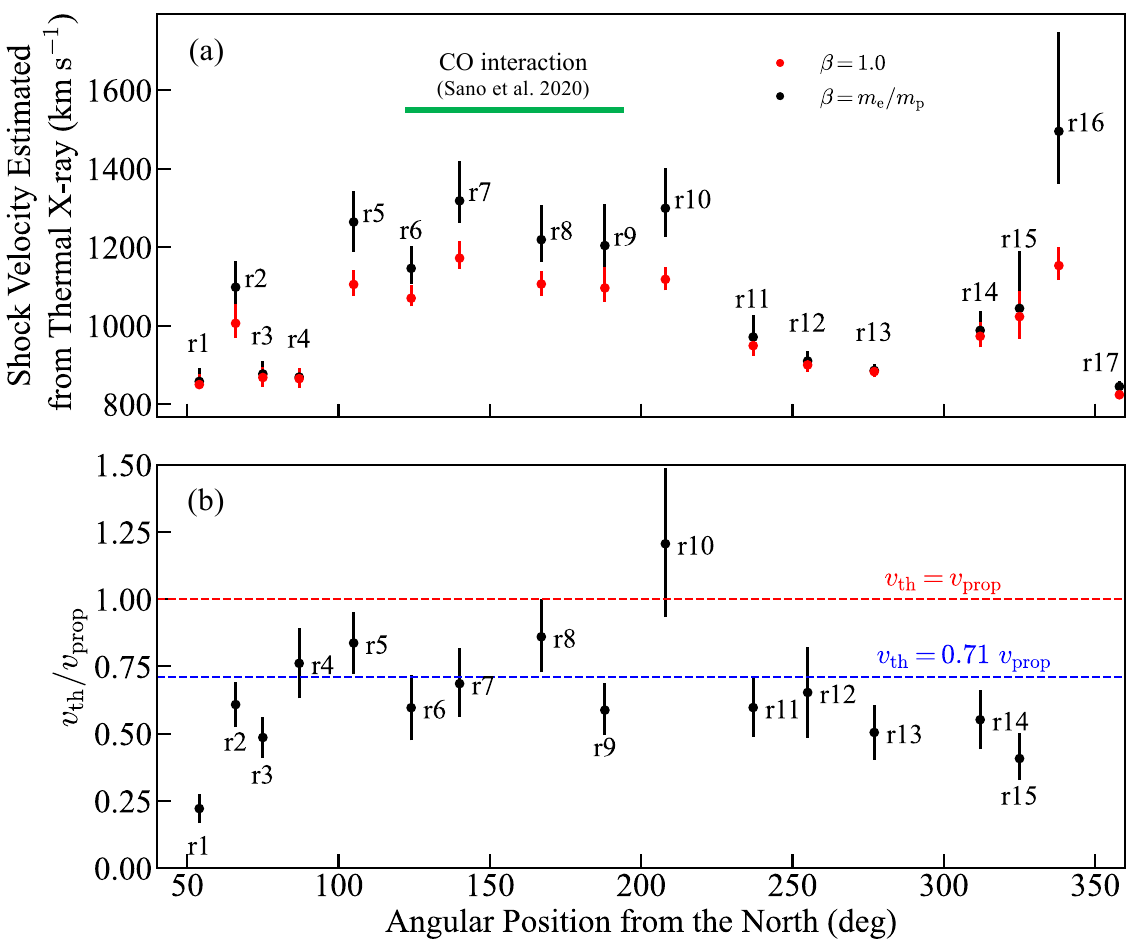} 
\end{center} 
\caption{\black{(a)} Spatial distribution of the shock velocities estimated from the thermal X-ray spectra extracted from regions along the rim of N132D. Horizontal axis represents angular position of each region (see Figure\ref{fig:region}) from north in a anticlockwise direction. The fitting results of shock velocity for $\beta$ = $m_{\rm e}$/$m_{\rm p}$ (red) and $\beta$ = 1.0 (black) are summarized. 
The green line shows the regions that overlap with the shock interaction with a CO cloud.
\black{(b)~The ratio of the shock velocities estimated from the thermal X-ray spectra ($v_{\rm th}$) and proper motion velocities ($v_{\rm prop}$; \citealt{2024AAS...24311312P}). The red and blue lines indicate $v_\mathrm{th} = v_\mathrm{prop}$ and $v_\mathrm{th} = 0.71\, v_\mathrm{prop}$, respectively. The blue line corresponds to the maximum discrepancy (lower limit of $v_\mathrm{th}/v_\mathrm{prop}$) achievable by a combination of the adiabatic cooling (Section~\ref{ss:Adiabatic Cooling}), shock obliquity, and magnetic field (Section~\ref{ss:MHD}).
Note that proper motion velocities are not available for the regions overlapping with r16 and r17 \citep{2024AAS...24311312P}.
}}
\label{fig:velocity}
\end{figure*}

We fit the spectra of the source regions with the model described above.
The best-fit model and spectrum of the region r1 are shown in Figure \ref{fig:a}.  
The spectra and the best-fit parameters of the other regions are given in Appendix.
Figure \ref{fig:velocity} \black{(a)} shows the spatial distribution of the shock velocity along the rim of N132D.
The shock velocities are found to have a significant variation in the range of 800--1500~$\rm{km~s^{-1}}$. 
To evaluate the effect of the collisionless electron heating, 
we consider the most efficient case with $\beta=1.0$.
The obtained shock velocity in this case is also shown in Figure \ref{fig:velocity} \black{(a)}.
The results with $\beta ={m_{\rm e}/m_{\rm p}}$ and $\beta=1.0$ correspond to the upper and lower limits of the estimated shock velocities, respectively.

\section{Discussion}\label{sec:discussion}
\black{We have estimated the forward shock velocity along the rim of N132D based on the thermal X-ray emission.}
Our results indicate that the forward shock velocity depends on the angular position, within the range of 800--1500~$\rm{km~s^{-1}}$.
\black{Such a significant spatial variation of shock velocities has been observed in other SNRs including SN1006 \citep{2014ApJ...781...65W} and Tycho \citep{2021ApJ...906L...3T}, and is believed to indicate a non-uniform environment or asymmetric explosion.
In this section, we discuss the environment and energetics of the shock in N132D by comparing our shock velocity estimates with other velocity measurements and multi-wavelength properties.}

In the southern rim (r5--r10), where the SNR exhibits the brightest emission from the shocked ISM, the shock velocities obtained are $\sim$\,1200~$\rm{km~s^{-1}}$. 
These values are consistent with the recent estimate by \cite{collaboration_xrism_2024},
where the forward shock velocity was estimated using the Doppler broadening of the Si and S He$\alpha$ lines from the entire SNR.
The shock radius ($R$) and velocity ($v_{\rm sh}$) derived from the Sedov-Taylor evolutionary model are, 
\begin{equation}
    R\approx8~ 
    \left(\frac{t}{2750~\mathrm{yr}}\right)^{\frac{2}{5} }
    \left(  \frac{E_0}{10^{51}\, \mathrm{erg}} \right) ^\frac{1}{5}
    \left(  \frac{n_0}{1\, \mathrm{cm}^{-3}} \right) ^{-\frac{1}{5}} 
    \mathrm{pc} 
\end{equation}
\begin{equation}\label{eq:Sedov}
    v_\mathrm{sh}\approx 1100\, 
    \left(\frac{t}{2750~\mathrm{yr}}\right)^{-\frac{3}{5} }
    \left(  \frac{E_0}{10^{51}\, \mathrm{erg}} \right) ^\frac{1}{5}
    \left(  \frac{n_0}{1~\mathrm{cm}^{-3}} \right) ^{-\frac{1}{5}}~\mathrm{km\,s^{-1}},
\end{equation}
where $t$, $E_0$, $n_0$ are the SNR age, kinetic energy released by the explosion, and ambient density, respectively. 
Our estimated velocity is similar to the one predicted by the Sedov model when we assume the age of N132D, the typical explosion energy and the typical number density of ISM.
This assumption roughly explains the radius of N132D ($\approx 10$~pc) as well.

\subsection{Discrepancy in Derived Shock Velocities}

Recent proper motion study has estimated the forward shock velocities using the Chandra data obtained in 2006 and 2019--2020 \citep{2024AAS...24311312P}.
Here we compare these measurements with our velocity estimates.
The observed proper motions, $\approx 0\farcs016~{\rm yr}^{-1}$ in the northern parts (r1, r2) and $\approx 0\farcs007~{\rm yr}^{-1}$ in the southern parts (r5--r13), correspond to the velocities of $v_{\rm prop}\approx3700$ $\rm{km~s^{-1}}$ and $v_{\rm prop}\approx1700~\rm{km~s^{-1}}$, respectively.
Therefore, the ratio between $v_{\rm th}$ (our measurement) and $v_{\rm prop}$ (the proper motion measurement) is systematically lower than unity in most regions \black{as summarized in Figure~\ref{fig:velocity} (b)}: 
$v_{\rm th}/v_{\rm prop} \approx 0.24$ (r1), 
$\approx 0.71$ (r5--r10), and $\approx 0.53$ (r11--r13). 
Such small $v_{\rm th}/v_{\rm prop}$ values are somewhat surprising because $v_{\rm th}$ reflects the past shock velocity when the plasma was shock-heated, while $v_{\rm prop}$ may be affected by recent deceleration due to an interaction with dense materials. 
Therefore, the discrepancy between $v_{\rm prop}$ and $v_{\rm th}$ implies that some physical processes have taken place in the shock transition regions to reduce the downstream thermal energy, or $v_{\rm th}$ is substantially underestimated with respect to the actual shock velocity. 
In the following subsections, we discuss possible processes responsible for the observed discrepancy. 

\subsubsection{Ionization in shock precursor}\label{ss:precursor}

Our estimate of the immediate post shock state might be affected by a possible bias in the preshock ionization state that can be modified by photons and/or charged particles in the shock precursor \citep{1979ApJ...227..131S, raymond_cygnus_2023}. 
In fact, the presence of low-charged ions such as O$^{2+}$ and S$^{1+}$ has been confirmed in the shock precursor of N132D \citep{1996AJ....112.2350M}. 
Although this effect is not taken into account in our \texttt{IONTENP} model, such slight ionization in the initial state does not affect significantly the thermal evolution of the downstream plasma. We conclude, therefore, that the ionization in the shock precursor is not a major cause of the observed discrepancy between $v_{\rm prop}$ and $v_{\rm th}$.

\subsubsection{Adiabatic Cooling}\label{ss:Adiabatic Cooling}

As a physical process that decreases the downstream thermal energy, we first consider an adiabatic expansion of the shock-heated plasma. 
In the adiabatic process of an ideal monoatomic gas, $TV^{\gamma-1}$ is conserved, where $T$ and $V$ are the gas volume and temperature, respectively.  
Assuming that the SNR expansion is spherically symmetric and $\gamma=5/3$, we can convert the relation $TV^{\gamma-1} = {\rm const.}$ to $TR^2 = {\rm const.}$, where $R$ is the radius of the sphere.
If we assume that the plasmas in our analysis regions have expanded while propagating from the inner to outer edges of the regions, plasma temperatures before and after the expansion ($T_1$ and $T_2$, respectively) can be related as,
\begin{equation}
    \frac{T_1}{T_2}=\left(\frac{R}{R-W}\right)^2,
\end{equation}
where $R$ and $W$ are the current radius of N132D and the width of the analyzed regions. Assuming typical values, $R=10~\rm pc$ and $W=0.7~\rm pc$, we obtain a temperature ratio of 1.1, indicating a $\approx14$\% decrease in temperature.
This corresponds to a velocity discrepancy of $v_{\rm th}/v_{\rm prop}\approx0.93$, which is much higher than the observed value.

\subsubsection{Shock obliquity and magnetic field} \label{ss:MHD}
\color{black}
Under conditions of an oblique shock, only the perpendicular component of the shock velocity relative to the shock front $v_{\rm sh,\perp}$ directly contributes to shock heating. An overestimation of the downstream temperature based on the proper motion velocity $v_{\rm prop}$ is possible because $v_{\rm sh,\perp}\leq v_{\rm prop}$ can be realized.
The velocity $v_{\rm sh, \perp}$ can be written as $v_{\rm sh,\perp}=\alpha^{0.5}v_{\rm prop}$,
where the $\alpha$ is a coefficient to describe the shock obliquity.
Since the typical range of the coefficient $\alpha$ 
is estimated to be $\approx$ 0.6--0.9 \citep{Shimoda_2015}, 
$v_{\rm th}/v_{\rm prop}$ higher than $\sim$\,0.77 can be explained by the oblique shock effect. 

Given that the energy density of magnetic field is important in the shock transition, the downstream temperature is reduced compared to that without magnetic field.
The conservation of energy flux at a shock can be expressed as
\begin{equation}
(\epsilon_{\rm 1}+\epsilon_{\rm B_1}) v_1 = \left(\epsilon_{\rm 2}+\epsilon_{\rm B_2}+\epsilon_{\rm th}\right)v_2,\label{eq:energy_flux}
\end{equation}
where $\epsilon$, $\epsilon_{\rm B}$, $\epsilon_{\rm th}$, and $v$ represent the kinetic, magnetic field, thermal energy densities, and fluid velocity, respectively.
The subscripts 1 and 2 refer to the upstream and downstream, respectively.
We neglect the upstream thermal energy density, considering the low temperatures of the ISM.
The energy densities can be written as
${\epsilon_{\rm 1}} 
=\frac{1}{2}\rho_1 v^2_1,\,{\epsilon_{\rm 2}} 
=\frac{1}{2}\rho_2 v^2_2$, ${\epsilon_{\rm th}} 
=\frac{\gamma}{\gamma-1}\frac{1}{\chi}\left(1-\frac{1}{\chi}\right)\rho_2v^2_2$,
$\epsilon_{\rm B_1} =\rm B_1^2/4\pi$ 
, and $\epsilon_{\rm B_2} =\rm B_2^2/4\pi=\chi^2 \epsilon_{\rm B_1}$,
where $\rho$ and $B$ indicate the mass density and magnetic field strength, respectively.
The energy densities of the magnetic field here are for a perpendicular shock, where the magnetic field affects the energetics most significantly.
The energy densities $\epsilon_1, \epsilon_2$ and $\epsilon_{\rm th}$ can be rewritten using $v_{\rm{prop}}$ (assumed to approximate $v_{\rm{sh}}$) and $v_{\rm{th}}$ as
\begin{equation}
    \begin{split}
        {\epsilon_{\rm 1}} 
&=\frac{1}{2}\rho_1 v^2_{\rm{prop}}\\
{\epsilon_{\rm 2}} 
&=\frac{1}{2}\rho_2 \left(\frac{1}{\chi}v_{\rm{prop}}\right)^2 \\
    \epsilon_{\rm th} &= \frac{\gamma}{\gamma-1} \frac{3}{16}\rho_2 v^2_{\rm th}.\label{eq:energy}
    \end{split}
\end{equation}
If $B_1=$ 20--100~$\rm \mu G$ \citep{2021A&A...655A...7H} and  $\chi=4$, the fraction of the energy flux transferred from the shock kinetic energy into the magnetic field is estimated to be $\approx3\%$ in r1 ($v_{\rm prop}=3700~\rm{km~s^{-1}}$) and $\approx17\%$ in r5--r10 ($v_{\rm prop}=1700~\rm{km~s^{-1}}$), corresponding to $v_{\rm th}/v_{\rm prop}\approx0.98$ and $\approx0.91$, respectively.

The discrepancies between $v_{\rm th}$ and $v_{\rm prop}$ in r5--r10 ($v_{\rm th}/v_{\rm prop}\approx$ 0.71) can thus be explained by a combination of the effects above: the adiabatic cooling, shock obliquity, and magnetic field discussed in Section \ref{ss:Adiabatic Cooling} and \ref{ss:MHD}.
However, it remains difficult to explain the larger discrepancies found in the other regions. 
\color{black}

\subsubsection{Particle Acceleration} \label{sec:CR}
The regions other than r5–r10 exhibit large discrepancies reaching a factor of $\approx4.3$ ($v_{\rm th}/v_{\rm prop}\approx0.23$).
Here we consider the contribution of particle acceleration to explain the discrepancies.
If SNRs are responsible for the observed energy density of cosmic rays, $\sim10\%$ of the kinetic energy of a supernova explosion has to be transferred to particle acceleration on average (e.g., \citealt{Vink_2011}).
We introduce an additional term accounting for the energy density of accelerated particles, $\epsilon_{\rm CR}$, on the right-hand side (downstream) of Equation~\ref{eq:energy_flux}.
The acceleration efficiency $\eta$ can be defined as the energy density of accelerated particles relative to the decrease in kinetic energy densities from upstream to downstream,
\begin{equation}
\begin{split}
 \eta=\frac{\epsilon_{\rm CR}}{\chi\epsilon_{\rm 1}-\epsilon_{\rm 2}}= \frac{\chi\epsilon_{\rm 1}-\epsilon_{\rm 2}-(\epsilon_{\rm th}+\epsilon_{\rm B_2}-\chi\epsilon_{\rm B_1})}{\chi\epsilon_{\rm 1}-\epsilon_{\rm 2}}.
\end{split}
\end{equation}
To explain $v_{\rm th}/v_{\rm prop}\approx0.23$ in region r1, where the discrepancy is the largest, the required acceleration efficiency would reach as high as $\eta\approx$ 90\% for both $\gamma=\frac{5}{3}$ and $\gamma=\frac{4}{3}$, corresponding to the two extreme cases where the downstream particles are either completely thermal or non-thermal (relativistic), respectively.

The resultant implication that N132D is an efficient accelerator is supported by $\gamma$-ray observations \citep{2021A&A...655A...7H}. The total energy of accelerated protons given by \cite{2021A&A...655A...7H} is $W_{\rm{p}}=4\times10^{50} \black{(n_\mathrm{p}/{10~\mathrm{cm^{-3}})^{-1}}}~\rm{erg}$, which can be translated to an acceleration efficiency of $\eta\sim40\%$ with a typical kinetic energy of the supernova explosion of $10^{51}$~erg.
Such high acceleration efficiencies would lead to enhanced synchrotron X-ray mission, which has not been detected in this SNR with the flux upper limit of $7.3\times10^{-13}~\rm erg~s^{-1}~cm^{-2}$ in the 2--10~keV band \citep{bamba_transition_2018}.
This level of the flux, however, does not immediately contradict the inferred acceleration efficiencies given the gamma-ray turnover at several TeV as already discussed in \cite{2021A&A...655A...7H}.
\black{Its high gamma-ray luminosity ($\sim 10^{36}$~erg~s$^{-1}$) favors the hadronic origin \citep{2021A&A...655A...7H} and thus the emission would mainly originate from the dense cloud interacting with the southern part of N132D \citep{sano_alma_2020}. 
The discrepancy in the southern region, $v_{\rm th}/v_{\rm prop} \approx 0.71$, corresponds to an acceleration efficiency of $\eta \lesssim 50\%$, which is consistent with the value derived in the hadronic interpretation. 
Although the northern part of N132D shows a blowout structure in several different wavelengths (see Section~\ref{sec:gas}) probably associated with high shock velocities, this region would have only a minor contribution to the observed gamma rays because the gas density is low \citep{sano_alma_2020}.
Future high-resolution and sensitivity gamma-ray observations are desired to resolve the acceleration environment in the north.}

\subsection{Comparison with Distribution of Dense Gas}\label{sec:gas}

In the southern region of N132D, ALMA CO observations \citep{sano_alma_2020} detected molecular clouds interacting with the shock front, which overlap our analysis regions \black{r6--r9} (Figure~\ref{fig:velocity} \black{(a)}).
In the north, a part of the X-ray-emitting shell is disrupted and located further away from the explosion center compared to the southern regions, suggesting that the shock propagated at a higher velocity than in the south \citep{chen_supernova_2003,2024AAS...24311312P}.
\black{This blowout structure has also been observed in infrared \citep{rho_infrared_2023} and optical \citep[H$\alpha$;][]{law_three-dimensional_2020} images, and \ion{H}{1} gas distribution \citep{kim_neutral_2003}.}
Nevertheless, our spectroscopy shows that the velocities $v_{\rm th}$ are higher on average in the south (r5--r10) than in the north (r1--r4 and r14--r17).
Thus, the spatial distribution of the shock velocities shows no clear correlations with the distribution of the molecular cloud or the X-ray shell structure (Figure \ref{fig:region}).

The absence of clear correlations between $v_{\rm th}$ and the cloud density can be explained from two different perspectives, depending on whether the actual shock velocities are close to $v_{\rm prop}$ or $v_{\rm th}$.
If we assume $v_{\rm sh} \approx v_{\rm prop}$ whereas $v_{\rm th}$ is affected by the efficient particle acceleration as discussed above, smaller $v_{\rm prop}$ in the south would be consistent with denser environment, considering a shock deceleration in the cloud. According to this interpretation, one can find that regions with higher shock velocities are likely associated with higher acceleration efficiencies (Section~\ref{sec:CR}), indicating that the shock velocity is governing the acceleration processes in N132D.

Alternatively, in the other case where $v_{\rm sh} \approx v_{\rm th}$, two scenarios are possible.
First, if the cloud in the south is clumpy, it allows the shock wave to propagate through lower-density intercloud regions without significant deceleration \citep{Inoue_2012,Sano_2021}.
Second, the estimated velocity $v_{\rm th}$ represents the velocity in the past when the plasma was heated. Since the estimated ionization timescales ($\tau \gtrsim \rm{10^{11}~s ~cm^{-3}}$; Table~\ref{tab:spectra}) would convert to $> 100$ yrs, a recent interaction with the cloud may not yet affect the observed thermal emission.
Both scenarios would be able to qualitatively explain higher $v_{\rm th}$ values in the south than in the north without taking into account high acceleration efficiencies considered in Sections~\ref{sec:CR}.
Further high spatial-resolution radio observations are desired to determine whether the interacting cloud is clumpy or uniform, which would provide insights into the acceleration environment of N132D.

\section{Conclusions}\label{sec:conclusion}

In this work, we estimated the shock velocities along the outermost rim of the supernova remnant N132D by tracing back the thermal equilibration and the ionization processes of the X-ray emitting plasma. The resultant velocities show an azimuthal dependence in the range of 800--1500~$\rm{km~s^{-1}}$, without clear correlations with the distribution of the molecular or atomic gas, or with its horseshoe morphology in X-rays. We compared the shock velocity estimated in our analysis with those based on the proper motion and Doppler broadening measurements. These velocities are roughly consistent with each other in the southern part. In the north, however, a discrepancy reaching a factor $\approx 4.3$ was found between our estimates and the proper motion measurements. This discrepancy cannot be explained by the effects of adiabatic cooling, shock obliquity, or magnetic field.
This may indicate that the downstream thermal energy has been significantly reduced compared with the thermal energy simply predicted by the Rankine-Hugoniot conditions, suggesting a highly efficient particle acceleration (reaching $\eta \approx 90\%$) operating at part of the shock front.

\begin{acknowledgments}
We are grateful to Professor Hidetoshi Sano, Dr. Jiro Shimoda and Dr. Tomohiko Oka for their valuable comments and suggestions on the interpretation of the observational results.
This work has been supported by Grants-in-Aid for Scientific Research (KAKENHI) of the Japanese Society for the Promotion of Science (JSPS) grant Nos.\ JP22KJ1047 (Y.Ohshiro), JP22KJ3059, JP24K17093 (H.S.), JP23K22522, JP23K25907, JP23H04899 (R.Y), JP22H00158, and JP23H01211 (H.Y.).
\black{This paper employs a list of Chandra datasets, obtained by the Chandra X-ray Observatory, contained in ~\dataset[DOI: 10.25574/cdc.344]{https://doi.org/10.25574/cdc.344}.}

\end{acknowledgments}
\vspace{5mm}
\facilities{Chandra (ACIS)}

\software{HEAsoft \citep{2014ascl.soft08004N},
    CIAO \citep{10.1117/12.671760}, XSPEC \citep{1996ASPC..101...17A}, CALDB \citep{2007ChNew..14...33G}.}

\bibliographystyle{aasjournal}
\vspace{-0.5cm}
\small
\bibliography{references}

\begin{figure*}[ht]
  \begin{center}
    \includegraphics[width=160mm]{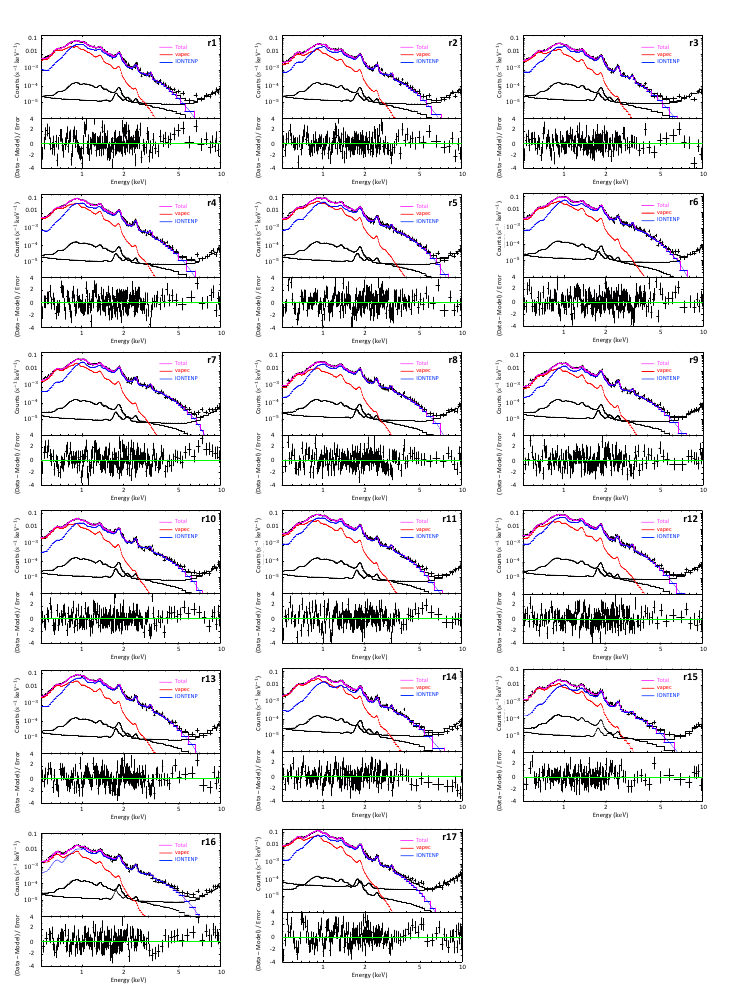}
  \end{center}
  \caption{Chandra spectra and best-fit models for all the analysis regions r1--r17. The magenta lines represent the entire models. The red and blue lines show the low- (\texttt{vpapec}) and high-temperature (\texttt{IONTENP}) plasma models, respectively. The two black lines indicate the detector and sky background.}
  \label{fig:spectra_all}
\end{figure*}

\pagebreak
\appendix
Table \ref{table:observation} shows the information of the Chandra observations used in our analysis. Table \ref{tab:sky} lists the best-fit values for the absorption column density in the LMC, source (\texttt{vapec}+\texttt{IONTENP}) model and C-stat/d.o.f.
All the spectra and the best-fit models for the analysis regions are summarized in Figure \ref{fig:spectra_all}.
\begin{table}[h]
    \centering
         \caption{Chandra Observation Logs of the N132D}
    {\begin{tabular}{c c c}
         \hline
        OBSID & Date & Exposure Time (ks) \\
        \hline \hline
        21362 & 2019-03-27 & 34.43 \\
        21363 & 2019-08-29 & 45.95 \\
        21364 & 2019-09-01 & 20.78 \\
        22687 & 2019-09-02 & 34.39 \\
        22094 & 2019-09-10 & 36.18 \\
        21687 & 2019-09-11 & 24.73 \\
        22841 & 2019-09-12 & 36.49 \\
        22853 & 2019-09-22 & 19.79 \\
        22740 & 2019-09-26 & 19.79 \\
        22858 & 2019-09-27 & 19.79 \\
        22859 & 2019-09-28 & 18.8 \\
        21881 & 2019-10-04 & 23.26 \\
        22860 & 2019-10-06 & 17.81 \\
        23270 & 2020-05-29 & 27.69 \\
        21882 & 2020-05-30 & 34.60 \\
        21883 & 2020-05-31 & 32.62 \\
        23044 & 2020-06-02 & 52.85 \\
        21886 & 2020-06-05 & 43.01 \\
        21365 & 2020-06-07 & 56.31 \\
        23277 & 2020-06-08 & 14.85 \\
        21884 & 2020-06-09 & 42.49 \\
        21887 & 2020-06-10 & 51.37 \\
        21885 & 2020-06-25 & 21.27 \\
        21888 & 2020-07-11 & 24.73 \\
        23286 & 2020-07-12 & 14.87 \\
        23303 & 2020-07-12 & 24.73 \\
        21361 & 2020-07-13 & 31.06 \\
        23317 & 2020-07-16 & 43.06 \\
        \hline
    \end{tabular}}
    \label{table:observation}
\end{table}

\begin{splitdeluxetable*}{l l l l B l l l l l l l l l l l}
\tablecaption{Best-fit spectral-model parameters for the rim regions}
\label{tab:spectra}
\tablehead{
\colhead{Reg.} & \colhead{Abs.} & \multicolumn{2}{c}{\texttt{vapec}} & \multicolumn{10}{c}{\texttt{IONTENP} (\(\beta = m_e/m_p\))} & \colhead{C-stat/dof} \\
\cline{3-4} \cline{5-14}
\colhead{} & \colhead{$N_{\rm H}$ (10$^{22}$ cm$^{-2}$)} & \colhead{$kT$ (keV)} & \colhead{${\rm Norm\tablenotemark{a}}$} & \colhead{${v_{\rm sh}}$ (km s$^{-1}$)} &\colhead{$kT\tablenotemark{b}$ (keV)} & \colhead{$\tau$ (10$^{11}$~cm$^{-3}$~s$^{-1}$)} & \colhead{O} & \colhead{Ne} & \colhead{Mg} & \colhead{Si} & \colhead{S} & \colhead{Fe} & \colhead{${\rm Norm\tablenotemark{c}}$} & \colhead{}
}
\startdata
r1 & $0.015_{-0.015}^{+0.040}$ & $0.27_{-0.01}^{+0.01}$ &$1.5_{-0.2}^{+0.6} $ & $858_{-15}^{+35}$ &$0.82_{-0.05}^{+0.07}$ &$2.07_{-0.43}^{+0.42}$ &$0.19_{-0.02}^{+0.02}$ &$0.23_{-0.02}^{+0.02}$ &$0.20_{-0.02}^{+0.02}$ &$0.26_{-0.02}^{+0.02}$ &$0.30_{-0.04}^{+0.04}$ &$0.20_{-0.02}^{+0.02}$ &$3.3_{-0.3}^{+0.3}$ & 437/377
\\
r2 & $0.007_{-0.007}^{+0.042}$ & $0.29_{-0.03}^{+0.01}$ &$0.75_{-0.09}^{+0.21} $ & $1098_{-71}^{+67}$ &$0.99_{-0.12}^{+0.14}$ &$0.91_{-0.15}^{+0.24}$ &$0.21_{-0.03}^{+0.06}$&$0.31_{-0.03}^{+0.03}$ &$0.17_{-0.02}^{+0.03}$ &$0.27_{-0.02}^{+0.02}$ &$0.42_{-0.06}^{+0.06}$ &$0.12_{-0.02}^{+0.04}$  &$1.2_{-0.2}^{+0.3}$ &346/367
\\
r3 & $0.017_{-0.17}^{+0.05}$ & $0.24_{-0.01}^{+0.01}$ &$1.0_{-0.2}^{+0.4} $ & $877_{-26}^{+34}$ &$0.82_{-0.06}^{+0.07}$ &$1.62_{-0.24}^{+0.32}$ &$0.24_{-0.04}^{+0.06}$ &$0.34_{-0.03}^{+0.05}$ &$0.27_{-0.03}^{+0.04}$ &$0.30_{-0.03}^{+0.03}$ &$0.46_{-0.06}^{+0.07}$ &$0.22_{-0.04}^{+0.05}$ &$2.1_{-0.2}^{+0.2}$ & 314/327
\\
r4 & $0.12_{-0.03}^{+0.03}$ & $0.25_{-0.03}^{+0.03}$ &$2.8_{-0.5}^{+0.6} $ & $869_{-17}^{+24}$ &$0.82_{-0.04}^{+0.05}$ &$1.78_{-0.24}^{+0.23}$ &$0.17_{-0.02}^{+0.03}$ &$0.29_{-0.02}^{+0.02}$ &$0.27_{-0.02}^{+0.02}$ &$0.29_{-0.02}^{+0.02}$ &$0.47_{-0.05}^{+0.05}$ &$0.21_{-0.03}^{+0.03}$ &$4.0_{-0.4}^{+0.3}$ & 409/363
\\
r5 & $0.18_{-0.03}^{+0.04}$ & $0.29_{-0.01}^{+0.01}$ &$3.3_{-0.5}^{+0.6} $ & $1264_{-76}^{+79}$ &$1.1_{-0.1}^{+0.1}$ &$0.83_{-0.08}^{+0.16}$ &$0.19_{-0.02}^{+0.03}$ &$0.29_{-0.02}^{+0.02}$ &$0.18_{-0.02}^{+0.02}$ &$0.26_{-0.02}^{+0.02}$ &$0.45_{-0.04}^{+0.05}$ &$0.14_{-0.01}^{+0.02}$ &$4.1_{-0.4}^{+0.5}$ & 538/405
\\
r6 & $0.18_{-0.03}^{+0.03}$ & $0.28_{-0.01}^{+0.01}$ &$2.8_{-0.4}^{+0.4} $ & $1146_{-39}^{+57}$ &$1.1_{-0.1}^{+0.1}$ &$1.24_{-0.11}^{+0.11}$ &$0.21_{-0.03}^{+0.04}$ &$0.32_{-0.03}^{+0.04}$ &$0.22_{-0.02}^{+0.02}$ &$0.30_{-0.02}^{+0.02}$ &$0.31_{-0.03}^{+0.03}$ &$0.16_{-0.02}^{+0.02}$ &$4.6_{-0.3}^{+0.4}$ & 502/420
\\
r7 & $0.19_{-0.01}^{+0.02}$ & $0.27_{-0.01}^{+0.01}$ &$1.6_{-0.2}^{+0.2} $ & $1318_{-56}^{+101}$ &$1.2_{-0.1}^{+0.1}$ &$0.97_{-0.11}^{+0.10}$ &$0.24_{-0.04}^{+0.06}$ &$0.36_{-0.04}^{+0.05}$ &$0.24_{-0.02}^{+0.03}$ &$0.32_{-0.03}^{+0.03}$ &$0.37_{-0.04}^{+0.05}$ &$0.15_{-0.03}^{+0.03}$ &$2.5_{-0.3}^{+0.2}$ & 456/417
\\
r8 & $0.41_{-0.11}^{+0.08}$ & $0.23_{-0.01}^{+0.03}$ &$1.9_{-0.7}^{+1.0} $ & $1219_{-56}^{+89}$ &$1.1_{-0.1}^{+0.1}$ &$0.96_{-0.13}^{+0.12}$ &$0.27_{-0.06}^{+0.08}$ &$0.33_{-0.06}^{+0.07}$ &$0.31_{-0.03}^{+0.03}$ &$0.33_{-0.03}^{+0.04}$ &$0.40_{-0.05}^{+0.05}$ &$0.29_{-0.07}^{+0.07}$ &$2.3_{-0.2}^{+0.2}$ & 377/400
\\
r9 & $0.25_{-0.07}^{+0.06}$ & $0.27_{-0.01}^{+0.02}$ &$0.89_{-0.23}^{+0.22} $ & $1204_{-72}^{+106}$ &$1.1_{-0.1}^{+0.2}$ &$1.03_{-0.14}^{+0.18}$ &$0.23_{-0.05}^{+0.08}$ &$0.37_{-0.05}^{+0.07}$ &$0.28_{-0.03}^{+0.04}$ &$0.36_{-0.03}^{+0.05}$ &$0.48_{-0.06}^{+0.08}$ &$0.20_{-0.04}^{+0.05}$ &$1.5_{-0.2}^{+0.1}$ & 350/388
\\
r10 & $0.24_{-0.07}^{+0.05}$ & $0.22_{-0.01}^{+0.02}$ &$2.7_{-0.9}^{+1.0} $ & $1299_{-73}^{+102}$  &$1.1_{-0.1}^{+0.1}$ &$0.73_{-0.10}^{+0.10}$ &$0.20_{-0.03}^{+0.04}$ &$0.27_{-0.03}^{+0.03}$ &$0.26_{-0.03}^{+0.03}$ &$0.32_{-0.03}^{+0.03}$ &$0.39_{-0.05}^{+0.06}$ &$0.26_{-0.05}^{+0.05}$ &$1.8_{-0.2}^{+0.1}$ &367/396
\\
r11 & $0.08_{-0.05}^{+0.05}$ & $0.26_{-0.01}^{+0.01}$ &$2.1_{-0.7}^{+0.7} $ & $971_{-28}^{+55}$ &$0.90_{-0.06}^{+0.10}$ &$1.20_{-0.17}^{+0.15}$ &$0.18_{-0.02}^{+0.03}$ &$0.24_{-0.02}^{+0.02}$ &$0.29_{-0.02}^{+0.02}$ &$0.29_{-0.02}^{+0.02}$ &$0.30_{-0.04}^{+0.04}$ &$0.23_{-0.02}^{+0.03}$ &$4.0_{-0.3}^{+0.3}$ & 408/371
\\
r12 & $0.07_{-0.04}^{+0.05}$ & $0.31_{-0.01}^{+0.01}$ &$1.1_{-0.2}^{+0.4} $ & $910_{-20}^{+25}$ &$0.88_{-0.04}^{+0.05}$ &$1.72_{-0.16}^{+0.19}$ &$0.29_{-0.04}^{+0.05}$ &$0.36_{-0.03}^{+0.03}$ &$0.33_{-0.02}^{+0.02}$ &$0.34_{-0.02}^{+0.02}$ &$0.38_{-0.04}^{+0.04}$ &$0.23_{-0.02}^{+0.03}$ &$4.3_{-0.3}^{+0.3}$ & 369/368
\\
r13 & $0.13_{-0.04}^{+0.04}$ & $0.25_{-0.01}^{+0.01}$ &$1.5_{-0.3}^{+0.3} $ & $886_{-13}^{+16}$ &$0.87_{-0.03}^{+0.04}$ &$2.13_{-0.17}^{+0.19}$ &$0.31_{-0.05}^{+0.06}$ &$0.41_{-0.03}^{+0.04}$ &$0.38_{-0.03}^{+0.03}$ &$0.35_{-0.02}^{+0.03}$ &$0.42_{-0.04}^{+0.04}$ &$0.28_{-0.03}^{+0.04}$ &$4.1_{-0.3}^{+0.2}$ & 414/361
\\
r14 & $0.14_{-0.03}^{+0.04}$ & $0.28_{-0.01}^{+0.01}$ &$1.6_{-0.2}^{+0.3} $ & $990_{-36}^{+51}$ &$1.0_{-0.1}^{+0.1}$ &$1.97_{-0.27}^{+0.30}$ &$0.25_{-0.04}^{+0.05}$ &$0.34_{-0.04}^{+0.04}$ &$0.28_{-0.03}^{+0.03}$ &$0.32_{-0.03}^{+0.03}$ &$0.44_{-0.06}^{+0.06}$ &$0.15_{-0.02}^{+0.03}$ &$1.7_{-0.2}^{+0.2}$ & 348/340
\\
r15 & $0.22_{-0.09}^{+0.07}$ & $0.25_{-0.01}^{+0.03}$ &$0.7_{-0.2}^{+0.3} $ & $1044_{-60}^{+145}$ &$1.0_{-0.1}^{+0.2}$ &$1.23_{-0.37}^{+0.26}$ &$0.37_{-0.08}^{+0.12}$ &$0.40_{-0.07}^{+0.09}$ &$0.32_{-0.06}^{+0.07}$ &$0.36_{-0.05}^{+0.07}$ &$0.56_{-0.09}^{+0.14}$ &$0.23_{-0.06}^{+0.07}$ &$0.8_{-0.2}^{+0.2}$ & 266/307
\\
r16 & $0.096_{-0.086}^{+0.107}$ & $0.22_{-0.02}^{+0.04}$ &$0.77_{-0.19}^{+0.65} $ &$1495_{-133}^{+252}$ &$1.2_{-0.2}^{+0.3}$ &$0.52_{-0.10}^{+0.14}$ &$0.19_{-0.04}^{+0.07}$ &$0.25_{-0.04}^{+0.05}$ &$0.20_{-0.03}^{+0.04}$ &$0.21_{-0.07}^{+0.08}$ &$0.39_{-0.09}^{+0.14}$ &$0.12_{-0.06}^{+0.07}$ &$0.9_{-0.2}^{+0.2}$ & 411/385
\\
r17 & $0.034_{-0.032}^{+0.027}$ & $0.22_{-0.03}^{+0.01}$ &$4.5_{-0.5}^{+0.8} $ & $845_{-27}^{+14}$ &$0.77_{-0.04}^{+0.04}$ & $1.49_{-0.09}^{+0.35}$ &$0.18_{-0.01}^{+0.02}$ &$0.30_{-0.02}^{+0.02}$ &$0.23_{-0.02}^{+0.02}$ &$0.26_{-0.02}^{+0.02}$ &$0.37_{-0.04}^{+0.04}$ &$0.19_{-0.02}^{+0.02}$ &$4.8_{-0.3}^{+0.2}$ & 535/486\\
\enddata
\tablenotetext{a} {Normalization is defined as $10^{-11}(4\pi D^2)^{-1}\int{n_{\rm e}n_{\rm H}dV}$, where $D$, $n_{\rm e}$, and $n_{\rm H}$ stand for distance (cm) and electron and hydrogen number densities ($\rm{cm^{-3}}$), respectively.}
\tablenotetext{b} {Electron temperature calculated with {\tt IONTENP} from the parameters $v_{\rm th}, \tau$, and $\beta$.}
\tablenotetext{c}{Normalization is defined as $10^{-10}(4\pi D^2)^{-1}\int{n_{\rm e}n_{\rm H}dV}$.}
\end{splitdeluxetable*}

\end{document}